\documentclass[11pt]{article}
\usepackage{amssymb,amsfonts,amsmath}
\usepackage{subfigure}
\usepackage{graphicx}
\usepackage{verbatim}
\usepackage{color}
\usepackage{hyperref}
\hypersetup{colorlinks}
\usepackage{physics}
\usepackage{dsfont}

\definecolor{darkred}{rgb}{1,0,0}
\definecolor{darkgreen}{rgb}{0,0.5,0}
\definecolor{darkblue}{rgb}{0,0,1}
\definecolor{orange}{rgb}{1,0.5,0}
\definecolor{green}{rgb}{0,1,0}
\definecolor{purple}{rgb}{.5,0,1}

\hypersetup{ colorlinks,
linkcolor=darkred,
filecolor=darkgreen,
urlcolor=darkblue,
citecolor=darkblue,
linktocpage=true }

\definecolor{markcolor}{rgb}{.25,0,1}

\definecolor{markcolor2}{rgb}{1,0,0}

\definecolor{markcolor3}{rgb}{0,1,0}

\def\hybrid{\topmargin 10pt    \oddsidemargin 0.1in 
        \headheight 0pt \headsep 0pt
        \textwidth 15.0cm      
        \textheight 20.0cm       
        \marginparwidth .875in
        \parskip 5pt plus 1pt   \jot = 1.5ex}

\hybrid

\catcode`\@=11

\def\marginnote#1{}
\newcount\hour
\newcount\minute
\newtoks\amorpm
\hour=\time\divide\hour by60 \minute=\time{\multiply\hour by60
\global\advance\minute by-\hour}
\edef\standardtime{{\ifnum\hour<12 \global\amorpm={am}%
        \else\global\amorpm={pm}\advance\hour by-12 \fi
        \ifnum\hour=0 \hour=12 \fi
        \number\hour:\ifnum\minute<10 0\fi\number\minute\the\amorpm}}
\edef\militarytime{\number\hour:\ifnum\minute<10 0\fi\number\minute}

\def\draftlabel#1{{\@bsphack\if@filesw {\let\thepage\relax
   \xdef\@gtempa{\write\@auxout{\string
      \newlabel{#1}{{\@currentlabel}{\thepage}}}}}\@gtempa
   \if@nobreak \ifvmode\nobreak\fi\fi\fi\@esphack}
        \gdef\@eqnlabel{#1}}
\def\@eqnlabel{}
\def\@vacuum{}
\def\draftmarginnote#1{\marginpar{\raggedright\scriptsize\tt#1}}

\def\draft{\oddsidemargin -.5truein
        \def\@oddfoot{\sl preliminary draft \hfil
        \rm\thepage\hfil\sl\today\quad\militarytime}
        \let\@evenfoot\@oddfoot \overfullrule 3pt
        \let\label=\draftlabel
        \let\marginnote=\draftmarginnote
   \def\@eqnnum{(\theequation)\rlap{\kern\marginparsep\tt\@eqnlabel}%
\global\let\@eqnlabel\@vacuum}  }

\def\draft2{
        \def\@oddfoot{\sl preliminary draft \hfil
        \rm\thepage\hfil\sl\today\quad\militarytime}
        \let\@evenfoot\@oddfoot \overfullrule 3pt
        \let\label=\draftlabel
        \let\marginnote=\draftmarginnote
   \def\@eqnnum{(\theequation)\rlap{\kern\marginparsep\tt\@eqnlabel}%
\global\let\@eqnlabel\@vacuum}  }

\def\preprint{\twocolumn\sloppy\flushbottom\parindent 2em
        \leftmargini 2em\leftmarginv .5em\leftmarginvi .5em
        \oddsidemargin -.5in    \evensidemargin -.5in
        \columnsep .4in \footheight 0pt
        \textwidth 10.in        \topmargin  -.4in
        \headheight 12pt \topskip .4in
        \textheight 6.9in \footskip 0pt
        \def\@oddhead{\thepage\hfil\addtocounter{page}{1}\thepage}
        \let\@evenhead\@oddhead \def\@oddfoot{} \def\@evenfoot{} }

\def\numberbysection{\@addtoreset{equation}{section}
        \def\theequation{\thesection.\arabic{equation}}}

\def\underline#1{\relax\ifmmode\@@underline#1\else
        $\@@underline{\hbox{#1}}$\relax\fi}

\def\titlepage{\@restonecolfalse\if@twocolumn\@restonecoltrue\onecolumn
     \else \newpage \fi \thispagestyle{empty}\c@page\z@
        \def\thefootnote{\fnsymbol{footnote}} }

\def\endtitlepage{\if@restonecol\twocolumn \else \newpage \fi
        \def\thefootnote{\arabic{footnote}}
        \setcounter{footnote}{0}}  

\catcode`@=12 \relax

\def\figcap{\section*{Figure Captions\markboth
        {FIGURECAPTIONS}{FIGURECAPTIONS}}\list
        {Figure \arabic{enumi}:\hfill}{\settowidth\labelwidth{Figure
999:}
        \leftmargin\labelwidth
        \advance\leftmargin\labelsep\usecounter{enumi}}}
 \relax
\def\tablecap{\section*{Table Captions\markboth
        {TABLECAPTIONS}{TABLECAPTIONS}}\list
        {Table \arabic{enumi}:\hfill}{\settowidth\labelwidth{Table
999:}
        \leftmargin\labelwidth
        \advance\leftmargin\labelsep\usecounter{enumi}}}
 \relax
\def\reflist{\section*{References\markboth
        {REFLIST}{REFLIST}}\list
        {[\arabic{enumi}]\hfill}{\settowidth\labelwidth{[999]}
        \leftmargin\labelwidth
        \advance\leftmargin\labelsep\usecounter{enumi}}}
 \relax
\makeatletter
\newcounter{pubctr}
\def\publist{\@ifnextchar[{\@publist}{\@@publist}}
\def\@publist[#1]{\list
        {[\arabic{pubctr}]\hfill}{\settowidth\labelwidth{[999]}
        \leftmargin\labelwidth
        \advance\leftmargin\labelsep
        \@nmbrlisttrue\def\@listctr{pubctr}
        \setcounter{pubctr}{#1}\addtocounter{pubctr}{-1}}}
\def\@@publist{\list
        {[\arabic{pubctr}]\hfill}{\settowidth\labelwidth{[999]}
        \leftmargin\labelwidth
        \advance\leftmargin\labelsep
        \@nmbrlisttrue\def\@listctr{pubctr}}}
 \relax
\makeatother

\def\be{\begin{equation}}
\def\ee{\end{equation}}
\def\ba{\begin{eqnarray}}
\def\ea{\end{eqnarray}}

\newcommand{\fr}[1]{\mathfrak{#1}}
\newcommand\bb{\mathbb }
\newcommand\msf{\mathsf }
\newcommand\ti{\widetilde }
\newcommand\wh{\widehat }

\def\k{\kappa}
\def\r{\rho}
\def\a{\alpha}

\def\b{\beta}

\def\g{\gamma}

\def\d{\delta}
\def\D{\Delta}

\def\m{\mu}

\def\l{\lambda}

\def\cI{{\cal I}}

\def\cN{{\cal N}}

\def\no{\noindent}

\def\IR{\relax{\rm I\kern-.18em R}}

\def\bse{\begin{small}\begin{equation*}}
\def\ese{\end{equation*}\end{small}}

\begin{document}


\renewcommand{\theequation}{\thesection.\arabic{equation}}
\csname @addtoreset\endcsname{equation}{section}

\newcommand{\eqn}[1]{(\ref{#1})}

\begin{titlepage}
\begin{center}
\hfill
\vskip 0.0cm


\vskip .5in

{\Large \bf Darboux-B\"acklund transformations, dressing $\&$ impurities in multi-component NLS}

\vskip 0.5in

{{\bf Panagiota Adamopoulou$^{a}$, Anastasia Doikou$^{a}$ and Georgios Papamikos$^{b}$ }} \vskip 0.2in

 \vskip 0.02in
{\footnotesize
$^{a}$Department of Mathematics, Heriot-Watt University,\\
Edinburgh EH14 4AS, United Kingdom}
\\[2mm]

{\footnotesize
$^{b}$Department of Mathematics and Statistics, University of Reading,\\ 
Reading RG6 6AX, United Kingdom}

\vskip .1cm


{\footnotesize {\tt E-mail: p.adamopoulou@hw.ac.uk, a.doikou@hw.ac.uk, g.papamikos@reading.ac.uk }}\\

\end{center}

\vskip 1.0in

\centerline{\bf Abstract}
We consider the discrete and continuous vector non-linear Schr\"odinger (NLS) model. We focus on the case where space-like local discontinuities are present, and we are primarily interested in the time evolution on the defect point. This in turn yields the time part of a typical Darboux-B\"acklund transformation. Within this spirit we then explicitly work out the generic B\"acklund transformation and the dressing associated to both discrete and continuous spectrum, i.e. the Darboux transformation is expressed in the matrix and integral representation respectively.
\no

\vfill

\end{titlepage}
\vfill \eject


\tableofcontents

\section{Introduction}

The non-linear Schr\"odinger equation (NLS) is one of the fundamental equations in mathematical physics with numerous applications, e.g. in the theory of non-linear optics and ocean waves (see e.g. \cite{ocean, optics}) to name a few. The NLS equation is an exactly solvable model, and has been integrated using the Inverse Scattering Transform (IST) \cite{Z-S nls, akns}, (see also \cite{FT, ablo-cla}). The first vector generalisation of NLS was introduced by Manakov \cite{Man}, while further generalisations of the model have been discovered (see for instance \cite{DegLomb1, DegLomb2, Dimakis} and references therein). An alternative to the IST method for constructing solutions of integrable equations is the dressing method, which was first presented by Zakharov and Shabat (ZS) \cite{Z-S} and further developed in \cite{Z-M}. The dressing formulation is based on the concept of Darboux Transformations (DTs) \cite{Matveev}, and it is this approach that we follow in Sections 4 and 5, where we present the dressing method for the vector NLS equation (vNLS). 

A relevant problem within this frame is the  interpretation of local space(time)-like integrable impurities as Darboux-B\"acklund transformations. It was first observed in \cite{corrigan} that classical defects in integrable $1+1$ integrable field theories may be seen as ``frozen'' Darboux-B\"acklund transformations (see also \cite{haku}--\cite{doikou15}). Then along this spirit the notion of {\it quasi B\"acklund} transformation as defect was introduced in \cite{doikou15} in both discrete and continuous integrable systems. In the present investigation we explore the quasi B\"acklund transformation i.e. the defect for the discrete and continuous vector NLS model. The $sl_2$ discrete NLS model was studied in \cite{avan-doikou}, whereas the continuous generalized NLS in \cite{doikou14}. Here we generalize the study of the discrete vector NLS, and then we also consider the continuous vector NLS model. In the continuous case we mainly focus on the time evolution associated to the defect, and inspired by this we give some generic expressions on the B\"acklund transformations and dressing. A brief discussion of a novel class of BTs that associate solitonic with anti-solitonic solutions is also presented.

More precisely, the outline of this paper is as follows: in Section 2 we present the discrete vector NLS model, after a brief review we study the model in the presence of a local defect in section 3. The associated integrals of motion and the corresponding time components of the Lax pairs are presented. In Section 4 we focus on the dressing and B\"acklund transformations (BT) for the continuous vector NLS equation. We treat both the focusing and defocusing cases simultaneously using an appropriate symmetry of the related Lax pair. Such symmetry groups, known as reduction groups, were first introduced in \cite{mikhToda, MikhRG, Mikh1981} and later developed in, e.g. \cite{lombRG}. In particular, we present the dressing transformation and give the general higher rank $1$-soliton solution, as well as the $n$-soliton solution as a ratio of determinants. Moreover, we obtain the B\"acklund transformation for the vector NLS model, which generalises the BT for the focusing and defocusing scalar NLS equation presented in \cite{boitipemp, chen}. We also briefly discuss the existence of a novel class of BTs, which essentially relates each field to its conjugate or in other words solitonic solutions to anti-solitonic ones. In Section 5 we provide a generic description of the ZS dressing and the Darboux-B\"acklund transforms as integral representations. The novel case of different spectral parameters associated to each field even in the case of ``one-soliton'' solution is discussed. This picture is more in tune with the quantum picture and the nested Bethe ansatz formulation. Both the discrete and continuous spectrum are discussed for the vNLS model.

\section{Discrete vector NLS}
Let us first focus our analysis on the discrete vector NLS model, generalizing essentially the results presented in \cite{avan-doikou}, where the $sl_2$ NLS model was studied in the presence of point-like defects. We shall focus mainly on the time evolution of the degrees of freedom associated to the defect obtained essentially as equations of motion of the system evaluated on the defect point.
We consider the following linear system \cite{FT}
\be
\begin{aligned}
\Psi_{j+1} &= \bb L_{j} \Psi_{j}  \\
\dv{t} \Psi_{j} & = \bb A_{j}  \Psi_{j} \,,
\end{aligned}
\ee
for an auxiliary function $\Psi$, with $(\bb L, \bb A)$ the Lax pair. Here  $j$ denotes the lattice site on a one-dimensional $N$-site periodic lattice. The compatibility condition of the above system of equations reads
\be 
\label{zcc}
\dv{t} \bb L_{j} = \bb A_{j+1} \bb L_{j} - \bb L_{j} \bb A_{j}\,,
\ee
and is equivalent to the discrete (differential-difference) equation at hand.

In the case of the discrete $\fr gl_{\cN}$ NLS model, the associated Lax operator $\bb L$ (acting on site $j$) is given by
\be \label{L1 vNLS}
\bb{L}_j (\l) =  (1+ \l + \sum_{k=1}^{\cN-1} x_j^{(k)} X_j^{(k)} ) e_{11} + \sum_{k=2}^{\cN} e_{kk} + \sum_{k=2}^{\cN} \left(  x_{j}^{(k-1)}e_{1k} + X_{j}^{(k-1)} e_{k1}\right)  \,,
\ee
where $\l$ is the spectral parameter and $e_{kl}$ are $\cN \times \cN$ matrices such that $(e_{kl})_{pq} = \d_{kp} \d_{lq}$.  The Lax operator \eqref{L1 vNLS} satisfies the quadratic algebra \cite{FT}
\be \label{algebra}
\pb \Big {\bb{L}_{a i} (\l_1)} {\bb{L}_{b j} (\l_2)} = \Big [r_{a b}(\l_1 - \l_2),\bb L_{a i} (\l_1) \bb L_{b j} (\l_2)\Big ] \d_{i j} \,,
\ee
where the indices $a\ ,b$ denote auxiliary spaces and $i,\ j$ denote sites on the one-dimensional lattice. The corresponding $r$-matrix is \cite{yang}
\be \label{r matrix}
r(\l) = \frac{\bb P}{\l} \quad \mbox{with} \quad \bb P = \sum_{i,j = 1}^{\cN} e_{i j} \otimes e_{j i} \,,
\ee
and satisfies the classical Yang-Baxter equation \cite{sts}. Relation \eqref{algebra} provides the following Poisson brackets between the dynamical variables $x^{(k)}$, $X^{(k)}$
\be \label{PB}
\pb \Big {x_{i}^{(k)}}{X_{j}^{(l)}} = - \d_{i j} \d_{k l} \,.
\ee

The monodromy matrix is defined as the product of $N$ Lax operators each acting on a site of the periodic lattice, in other words,
\be \label{monodromy}
\bb T (\l) = \bb L_{N}(\l)\, \bb L_{N-1}(\l) \ldots \bb L_{1} (\l) \,.
\ee
The monodromy matrix $\bb T(\l)$ satisfies the same quadratic relation \eqref{algebra} as $\bb L(\l)$. If we define the transfer matrix $\tau(\l)$ as the trace of the monodromy matrix, i.e. $\tau (\l) = \tr \bb T (\l)$, then one can verify that  
\be 
\pb  \Big  {\tau(\l_1)}{\tau(\l_2)} = 0 \,.
\ee
Hence, expansion of the transfer matrix $\tau(\l)$ in powers of the spectral parameter $\l$ provides the charges in involution. To obtain the local integrals of motion, expansion of $\ln \tau (\l)$ is required instead.

In order to obtain the associated integrals of motion it is convenient to  utilise the bra-ket notation for a $(\cN - 1)$-dimensional vector and co-vector, in other words,
\be
\ket {X} :=
\begin{pmatrix}
X^{(1)}\\
X^{(2)}\\
\vdots \\
X^{(\cN-1)}
\end{pmatrix} ,
 \qquad 
\bra {x} := \left( x^{(1)} \: x^{(2)}  \dots \: x^{(\cN-1)}  \right) \,,
\ee
and also define 
\be
 \msf N_j = 1+  \sum_{k=1}^{\cN-1} x_j^{(k)} X_j^{(k)} := 1 + \braket{x_j}{X_j} \,.
\ee
Then the Lax operator \eqref{L1 vNLS} can be written in block matrix form as
\be\label{L vNLS}
\bb L_{j} = \l D_{j} + A_{j}  = 
\l \begin{pmatrix}
1 & \bra 0 \\
\ket 0 & \bf 0 
\end{pmatrix}
+
\begin{pmatrix}
\msf N_{j} & \bra{x_j} \\
\ket{X_j} & \mathds{1}
\end{pmatrix} ,
\ee
where $\bf 0$ and $\mathds{1}$ denote the $(\cN - 1) \times (\cN -1)$ zero and identity matrix, respectively. 

Expanding the monodromy matrix $\bb T(\l)$ in powers of $\frac{1}{\l}$ we can then express the transfer matrix in the form
\be
\tau (\l) = 1 +  \frac{1}{\l} \, \tau_1 + \frac{1}{\l^2} \, \tau_2 + \frac{1}{\l^3} \, \tau_3 + \ldots \,,
\ee
where for instance, 
\begin{align}
\tau_1 &= \tr \left( \sum_{i=1}^N D_N \ldots D_{i+1} A_i D_{i-1} \ldots D_1 \right) \nonumber \,, \\
\tau_2 &= \tr \left(  \sum_{i>j} D_N \ldots D_{i+1} A_i D_{i-1} \ldots D_{j+1} A_j D_{j-1} \ldots D_1  \right) \,.
\end{align}
The associated local integrals of motion are obtained as coefficients of the expansion of $\ln \tau (\l)$ in powers of $\frac{1}{\l}$, i.e.:
\be
\ln \tau (\l \rightarrow \infty) = \frac{1}{\l} \, \rm I_1 + \frac{1}{\l^2} \, \rm I_2 + \frac{1}{\l^3} \, \rm I_3 + \ldots \,.
\ee
The different $\rm I_m$  are found in terms of $\lbrace \tau_i \rbrace_{i=1}^m$. For example, for the first three local integrals of motion we have 
\be
\rm I_1 = \tau_1 \,, \quad \rm I_2 = - \frac{1}{2} \rm I_1^2 +  \tau_2  \,, \quad \rm I_3 = - \frac{1}{6} \rm I_1^{3} - \rm I_1 \rm I_2 + \tau_3 \,,
\ee
and so on. It turns out that the first three $\rm I_m$ are given by the following expressions
\begin{align}\label{IOM}
\rm I_1 &= \sum_{i=1}^N  \msf N_{i}  \,, \nonumber \\
\rm I_2 &= -\frac{1}{2}  \sum_{i=1}^N  \msf N_{i} ^2  +  \sum_{i=1}^N  \braket{x_{i}}{X_{i-1}}  \,, \nonumber \\
\rm I_3 &= \frac{1}{3}  \sum_{i=1}^N  \msf N_{i} ^3  +  \sum_{i=1}^N  \braket{x_{i}}{X_{i-2}} - \sum_{i=1}^N (\msf N_{i-1} + \msf N_{i}) \braket{x_{i}}{X_{i-1}} \,.
\end{align}

Each of the above integrals of motion has an associated Lax pair $(\bb L, \bb A)$. The operator $\bb A$ of the Lax pair can be found via \cite{sts}
\be
\bb A_j(\l,\m) = \tau^{-1}(\l)\, \tr_a \Big( \bb T_a (N,j, \l) \, r_{ab} (\l - \m) \, \bb T_a (j-1,1,\l) \Big) \,,
\ee
where we have defined 
\be \label{T_ij}
\bb T_a (i, j, \l) = \bb L_{a i} (\l) \, \bb L_{a i-1} (\l) \ldots \bb L_{a j} (\l) \,, \quad \mbox{with} \quad i>j \,.
\ee
In the present case, where the $r$-matrix is given by expression \eqref{r matrix},  operator $\bb A$ takes the form:
\be \label{A vNLS}
\bb A_j(\l,\m) = \frac{\tau^{-1} (\l)}{\l - \m} \, \bb T(j-1,1, \l) \, \bb T(N,j,\l) \,.
\ee
Expanding \eqref{A vNLS} in powers of $\frac{1}{\l}$ results in
\be
\bb A_{j} (\l, \m) =  \frac{1}{\l} \, \bb A^{(1)}_{j}(\m) + \frac{1}{\l^2} \, \bb A^{(2)}_{j}(\m) + \frac{1}{\l^3} \, \bb A^{(3)}_{j}(\m) + \ldots \,,
\ee
with each $A_{j}^{(i)}$ being associated to each of the integrals of motion \eqref{IOM}. In the present case we obtain 
\begin{align} \label{A123 vNLS}
\bb A_{j}^{(1)}(\m) &=  
\begin{pmatrix}
1 & \bra 0 \\
\ket 0 & \bf 0 
\end{pmatrix} \,, \qquad 
\bb A_{j}^{(2)}(\m) =  
\begin{pmatrix}
\m & \bra{x_j} \\
\ket{X_{j-1}} & \bf 0 
\end{pmatrix} \,, \nonumber \\
\bb A_{j}^{(3)}(\m) &=  
\begin{pmatrix}
\m^2 - \braket{x_j}{X_{j-1}} & \m \bra{x_j}  - \msf N_j \bra{x_j} + \bra{x_{j+1}} \\
\ket{X_{j-1}}\m  -  \ket{X_{j-1}}\msf N_{j-1} + \ket{X_{j-2}} & \ket{X_{j-1}}\bra{x_{j}} 
\end{pmatrix} \,.
\end{align}

Consider the pair $(\bb L_j, \bb A^{(3)}_j )$. Then the associated equations of motion for the multicomponent fields $\bra{x}$ and $\ket*{X}$ are obtained from the compatibility condition \eqref{zcc} 
\begin{align}\label{EOM}
\bra{\dot{x}_{j}} &=    \msf N_{j}^{2} \bra{x_j}  - \braket{x_{j+1}}{X_j} \bra{x_j} - \braket{x_j}{X_{j-1}} \bra{x_j} - (\msf N_j + \msf N_{j+1}) \bra{x_{j+1}} + \bra{x_{j+2}} \nonumber \,, \\
\ket*{\dot{X}_j} &=   - \ket*{X_j} \msf N_{j}^{2}  +  \ket*{X_j} \braket{x_{j+1}}{X_j} + \ket*{X_j} \braket{x_{j}}{X_{j-1}} + \ket*{X_{j-1}}(\msf N_{j-1} + \msf N_{j}) - \ket*{X_{j-2}} \,.
\end{align}


\section{Vector DNLS in the presence of defects}
We now consider the DNLS model in the presence of a point-like integrable defect. We introduce the defect on site $n$ of the one-dimensional $N$-site lattice, with $n \neq 1, N$. The Lax operator associated to the defect is
\be\label{L1 def}
\ti{\bb L}_n (\l) = \l \sum_{k=1}^{\cN} e_{kk} + \sum_{k,\,l = 1}^{\cN} \a^{(kl)}_n e_{kl} \,,
\ee
and we assume that it satisfies the same quadratic algebra \eqref{algebra} as the $\mathfrak{gl}_{\cN}$ Lax operator \eqref{L1 vNLS}. Hence, it follows that the Poisson brackets between the dynamical variables $\a^{(kl)}_n$ associated to the defect are given by
\be\label{PB def}
\pb{\a^{(ij)}_n}{\a^{(kl)}_n} = \a^{(il)}_n \d_{kj} - \a^{(lj)}_n \d_{ik} \,.
\ee
For convenience we write the defect Lax operator \eqref{L1 def} in the following form
\be\label{L def}
\ti {\bb L}_n (\l) = \l \mathds{1}_{\cN} + \ti{A}_n = \l \mathds{1}_{\cN} + 
\begin{pmatrix}
\a_n & \bra{\b_n} \\
\ket{\g_n} & {\bf \D}_n
\end{pmatrix} ,
\ee
where 
$\mathds{1}_{\cN}$ denotes the $\cN \times \cN$ identity matrix and 
\begin{align}
\a_n := \a_n^{(11)} \,, \quad \bra{\b_n} :=  \left( \a_n^{(12)} \: \a_n^{(13)}  \dots \: \a_n^{(1\cN)}  \right) \,, \nonumber \\
\ket{\g_n} := 
\begin{pmatrix}
\a_n^{(21)}\\
\a_n^{(31)}\\
\vdots \\
\a_n^{(\cN 1)}
\end{pmatrix} \,, \quad
{\bf \D}_n := 
\begin{pmatrix}
\a_n^{(22)} & \ldots & \a_n^{(2 \cN))} \\
\vdots & \ddots & \vdots \\ 
\a_n^{(\cN 2)} & \ldots & \a_n^{(\cN \cN)}
\end{pmatrix} \,.
\end{align}

In the case where a defect is introduced on site $n$, the monodromy matrix $\bb T(\l)$ reads
\be \label{monodromy def}
\bb T (\l) = \bb L_{N}(\l) \ldots   \bb L_{n+1}(\l)\ \ti{\bb L}_{n} (\l)\ \bb L_{n-1}(\l)  \ldots \bb L_{1} (\l) \,.
\ee
To obtain the local integrals of motion we first expand the monodromy matrix in powers of $\frac{1}{\l}$, where this time the contribution from the defect point must be taken into account (see also \cite{doikou14} for generic expressions). Then, the expansion in powers of $\frac{1}{\l}$ of the logarithm of the transfer matrix (with the defect incorporated) reads
\be
\ln \tau (\l \rightarrow \infty) = \frac{1}{\l} \, \cI_1 + \frac{1}{\l^2} \, \cI_2 + \frac{1}{\l^3} \, \cI_3 + \ldots \,.
\ee
The integrals of motion take the form
\begin{align}\label{IOM2}
\cI_1 &= \sum_{i \neq n}  \msf N_{i} + \a_n   \,, \nonumber \\
\cI_2 &= -\frac{1}{2}  \sum_{i \neq n} \msf N_{i} ^2 - \frac{1}{2} \a_n^{2} +  \sum_{i \neq n, n-1 }  \braket{x_{i+1}}{X_{i}} + \braket{x_{n+1}}{X_{n-1}} + \braket{\b_{n}}{X_{n-1}} + \braket{x_{n+1}}{\g_{n}}    \,, \nonumber \\
\cI_3 &= \frac{1}{3}  \sum_{i \neq n}  \msf N_{i} ^3 + \frac{1}{3}\a_n^{3}  +  \sum_{i \neq n, n \pm 1}  \braket{x_{i+1}}{X_{i-1}} - \sum_{i \neq n, n-1} (\msf N_{i} + \msf N_{i+1}) \braket{x_{i+1}}{X_{i}} + \msf N_{n+1} \braket*{ x_{n+1}}{\ti X_{n-1}}  \nonumber \\
&+ (\msf N_{n-1} - \a_n)\braket{\ti x_{n+1}}{X_{n-1}}  + \mel{x_{n+1}}{{\bf \D}_n}{X_{n-1}} + \braket*{x_{n+2}}{\ti X_{n-1}} + \braket{\ti x_{n+1}}{X_{n-2}} - \a_n \braket{x_{n+1}}{\g_n} \,.
\end{align}
where we have defined
\be
\bra{\ti x _{n+1}} = \bra{x_{n+1} + \b_n} \,, \qquad \ket*{\ti X_{n-1}} = \ket{X_{n-1} + \g_n} \,.
\ee
The components of the Lax pair around the defect point are given by:
\be
\bb A_n^{(2)} (\m) = 
\begin{pmatrix}
\m & \bra{\ti x_{n+1}} \\
\ket{X_{n-1}} & {\bf 0}
\end{pmatrix} \,, \qquad
A_{n+1}^{(2)} (\m) = 
\begin{pmatrix}
\m & \bra{ x_{n+1}} \\
\ket*{\ti X_{n-1}} & {\bf 0}
\end{pmatrix} \,.
\ee
\begin{align} \label{A3 def}
\bb A_{n-1}^{(3)}(\m) &=  
\begin{pmatrix}
\m^2 - \braket{x_{n-1}}{X_{n-2}} & (\m - \msf N_{n-1}) \bra{x_{n-1}} + \bra{\ti x_{n+1}} \\
\ket{X_{n-2}} (\m - \msf N_{n-2}) + \ket{X_{n-3}} & \ketbra{X_{n-2}}{x_{n-1}}
\end{pmatrix} \,, \nonumber \\
\bb A_{n}^{(3)}(\m) &=  
\begin{pmatrix}
\m^2 - \braket{\ti x_{n+1}}{X_{n-1}} &  \bra{\ti x_{n+1}} \mu  + \bra{\wh{x}_{n+1}} + \bra{x_{n+2}} \\
\ket{X_{n-1}} (\m - \msf N_{n-1}) + \ket{X_{n-2}} & \ketbra{X_{n-1}}{\ti x_{n+1}} 
\end{pmatrix} \,, \nonumber \\
\bb A_{n+1}^{(3)}(\m) &=  
\begin{pmatrix}
\m^2 - \braket*{x_{n+1}}{\ti X_{n-1}} & ( \m - \msf N_{n+1})\bra{x_{n+1}} + \bra{x_{n+2}} \\
 \ket*{\ti X_{n-1}} \m + \ket*{\wh X_{n-1}} + \ket{X_{n-2}} & \ketbra*{\ti X_{n-1}}{x_{n+1}} 
\end{pmatrix} \,,\nonumber \\
\bb A_{n+2}^{(3)}(\m) &=  
\begin{pmatrix}
\m^2 - \braket{x_{n+2}}{X_{n+1}} & (\m - \msf N_{n+2})\bra{x_{n+2}}  + \bra{x_{n+3}} \\
\ket{X_{n+1}} (\m  - \msf N_{n+1})  + \ket*{\ti X_{n-1}} & \ketbra{X_{n+1}}{x_{n+2}} 
\end{pmatrix} \,,
\end{align}
where
\ba
\bra{\wh{x}_{n+1}} = \bra{x_{n+1}}{{\bf \D}_n} - \msf N_{n+1} \bra{x_{n+1}}   -  \a_n \bra{\ti x_{n+1}} \,, \nonumber \\
\ket*{\wh X_{n-1}} = {{\bf \D}_n} \ket{X_{n-1}} -  \ket{X_{n-1}}\msf N_{n-1} - \ket*{\ti X_{n-1}} \a_n  \,.
\ea

Again we consider the pair $(\bb L_j, \bb A^{(3)}_j )$.  For $j \neq n, n \pm 1, n \pm 2$ the matrix $\bb A^{(3)}_j$ is given in \eqref{A123 vNLS} and the equations of motion for the fields $\bra{x}$, $\ket{X}$ coincide with equations \eqref{EOM}. However, in order to derive the equations of motion for the fields in the neighbourhood of the defect, i.e. for $j = n \pm 1, n \pm 2$, one must take into account expressions \eqref{A3 def}. Hence, we obtain the following differential-difference equations for $\bra{x}$, $\ket{X}$
\begin{align}\label{EOM -2 def}
\bra{\dot{x}_{n-2}} &= \msf N_{n-2}^2 \bra{x_{n-2}} - \braket{x_{n-1}}{X_{n-2}} \bra{x_{n-2}}  - \braket{x_{n-2}}{X_{n-3}} \bra{x_{n-2}} - (\msf N_{n-1} + \msf N_{n-2}) \bra{x_{n-1}}  \nonumber  \\
&+ \bra{\b_n} + \bra{x_{n+1}} \nonumber \,, \\
\ket*{\dot{X}_{n-2}} &= - \ket{X_{n-2}} \msf N_{n-2}^2  + \ket{X_{n-2}} \braket{x_{n-1}}{X_{n-2}}  + \ket{X_{n-2}} \braket{x_{n-2}}{X_{n-3}} +  \ket{X_{n-3}} (\msf N_{n-2} + \msf N_{n-3})\nonumber \\
&- \ket{X_{n-4}} \,,
\end{align}
\begin{align}\label{EOM -1 def}
\bra{\dot{x}_{n-1}} &= \msf N_{n-1}^2 \bra{x_{n-1}} - \braket{x_{n+1}}{X_{n-1}} \bra{x_{n-1}}  - \braket{x_{n-1}}{X_{n-2}} \bra{x_{n-1}} - \braket{\b_n}{X_{n-1}}\bra{x_{n-1}} \nonumber  \\
&- \a_n \bra{\b_n}  -  \msf N_{n-1} \bra{\b_n}- (\msf N_{n-1} + \msf N_{n+1}) \bra{x_{n+1}} + \bra{x_{n+1}} {\bf \D}_n + \bra{x_{n+2}}
\nonumber \,, \\
\ket*{\dot{X}_{n-1}} &= - \ket{X_{n-1}}  \msf N_{n-1}^2 + \ket{X_{n-1}} \braket{x_{n+1}}{X_{n-1}}  + \ket{X_{n-1}} \braket{x_{n-1}}{X_{n-2}}  + \ket{X_{n-1}} \braket{\b_n}{X_{n-1}} \nonumber \\
&+  \ket{X_{n-2}} (\msf N_{n-1} + \msf N_{n-2}) - \ket{X_{n-3}} \,,
\end{align}
\begin{align}\label{EOM +1 def}
\bra{\dot{x}_{n+1}} &= \msf N_{n+1}^2 \bra{x_{n+1}} - \braket{x_{n+1}}{X_{n-1}} \bra{x_{n+1}}  - \braket{x_{n+2}}{X_{n+1}} \bra{x_{n+1}} - \braket{x_{n+1}}{\g_n}\bra{x_{n+1}} \nonumber  \\
&- (\msf N_{n+1} + \msf N_{n+2}) \bra{x_{n+2}}  + \bra{x_{n+3}}
\nonumber \,, \\
\ket*{\dot{X}_{n+1}} &= -  \ket{X_{n+1}}\msf N_{n+1}^2 + \ket{X_{n+1}} \braket{x_{n+1}}{X_{n-1}}  + \ket{X_{n+1}} \braket{x_{n+2}}{X_{n+1}}  + \ket{X_{n+1}} \braket{x_{n+1}}{\g_n} \nonumber \\
&+ \ket{\g_n}\msf N_{n+1}  +  \ket{\g_n}\a_n +  \ket{X_{n-1}}\a_n - {\bf \D}_n \ket{X_{n-1}} +  \ket{X_{n-1}} (\msf N_{n-1} + \msf N_{n+1}) - \ket{X_{n-2}} \,,
\end{align}
\begin{align}\label{EOM +2 def}
\bra{\dot{x}_{n+2}} &= \msf N_{n+2}^2 \bra{x_{n+2}} - \braket{x_{n+2}}{X_{n+1}} \bra{x_{n+2}}  - \braket{x_{n+3}}{X_{n+2}} \bra{x_{n+2}} - (\msf N_{n+2} + \msf N_{n+3}) \bra{x_{n+3}} \nonumber  \\
&+ \bra{x_{n+4}}
\nonumber \,, \\
\ket*{\dot{X}_{n+2}} &= - \ket{X_{n+2}}\msf N_{n+2}^2  + \ket{X_{n+2}} \braket{x_{n+2}}{X_{n+1}}  + \ket{X_{n+2}} \braket{x_{n+3}}{X_{n+2}} +  \ket{X_{n+1}} (\msf N_{n+1} + \msf N_{n+2}) \nonumber \\
&- \ket{\g_n} - \ket{X_{n-1}} \,.
\end{align}
Moreover, on the defect site $n$ the compatibility condition \eqref{zcc} takes the form
\be
\dv{t} \ti {\bb  L}_n = \bb A_{n+1} \ti {\bb  L}_n -  \ti {\bb  L}_n \bb A_{n}\,,
\ee
and leads to the following equations of motion for the dynamical variables $\a_n , \bra{\b_n}, \ket{\g_n} , {\bf \D}_n$ associated to the defect
\begin{align}
\dot{\a}_n &= (\a_n + {\msf N_{n-1}}) \braket{\b_n}{X_{n-1}}  -  (\a_n + {\msf N_{n+1}} )  \braket{x_{n+1}}{\g_n}  + \braket{x_{n+2}}{\g_n} - \braket{\b_n}{X_{n-2}} \,, \nonumber\\
\bra*{\dot{\b_n}} &= - \braket{x_{n+1}}{X_{n-1}} \bra{\b_n} - \braket{x_{n+1}}{\g_n} \bra{\b_n} - \braket{\b_n}{X_{n-1}} \bra{\b_n} + \a^2_{n} \bra{\b_n} \nonumber \\
&+ \a_{n} (\a_n + {\msf N_{n+1}} )\bra{x_{n+1}}  -(\a_n + {\msf N_{n+1}} ) \bra{x_{n+1}} {\bf \D}_n -  \braket{\b_n}{X_{n-1}} \bra{x_{n+1}}   \nonumber \\
& - \a_n \bra{x_{n+2}} + \bra{x_{n+2}} {\bf \D}_n \,, \nonumber \\
\ket{\dot{\g_n}} &= - \ket{\g_n} \a_n^2 + \ket{\g_n} \braket{x_{n+1}}{\g_n}  + \ket{\g_n} \braket{x_{n+1}}{X_{n-1}} + \ket{\g_n} \braket{\b_n}{X_{n-1}} \nonumber \\
& - \ket{X_{n-1}} \a_n^2 - \ket{X_{n-1}} \a_n \msf N_{n-1} + {\bf \D}_n \ket{X_{n-1}} (\a_n + \msf N_{n-1}) + \ket{X_{n-1}} \braket{x_{n+1}}{\g_n} \nonumber \\
& +  \ket{X_{n-2}} \a_n  -  {\bf \D}_n \ket{X_{n-2}} \,, \nonumber \\
{\dot {\bf \D }}_n & = - (\a_n + {\msf N_{n-1}}) \ketbra{X_{n-1}}{\b_n}  + (\a_n + {\msf N_{n+1}}) \ketbra{\g_n}{x_{n+1}} + \ketbra{X_{n-2}}{\b_n} - \ketbra{\g_n}{x_{n+2}} \nonumber \\
& + \ketbra{X_{n-1}}{x_{n+1}} {\bf \D}_n - {\bf \D}_n \ketbra{X_{n-1}}{x_{n+1}} \,.
\end{align}
The equations above generalize the results presented in \cite{doikou14} in the $gl_{\cal N}$ case.

\section{The continuous vector NLS}
\label{sec:hier}
In the present section we construct a Darboux transformation for the continuous vNLS 
equations for both the focusing and defocusing cases. The construction makes use of the reduction group \cite{mikhToda}. The matrix that defines the Darboux transformation, is structurally  similar to the defect matrix as already discussed. This suggests an alternative to the $r$-matrix construction of the defect and possible classification using the theory of reduction groups.

We start with the Lax operator of the vector AKNS hierarchy namely,
\begin{equation}
\mathcal{L}(\lambda)=D_x-\mathbb{U}(\lambda), \quad \mathbb{U}(\lambda)=\lambda \mathbb{U}_1+\mathbb{U}_0
\label{eq:U-AKNS}
\end{equation}
where 
\begin{equation}
\mathbb{U}_1=\alpha\left(\begin{array}{cc}
\rho  \mathds{1} & 0 \\
0 & -1
\end{array}\right)
\label{eq:Us} \quad \text{and} \quad \mathbb{U}_0=\beta\left(\begin{array}{cc}
0 & \ket{v} \\
\bra{u} & 0
\end{array}\right).
\end{equation}
In \eqref{eq:Us} $\alpha$ and $\beta$ are complex numbers, $\mathds{1}$ is the $(\mathcal{N}-1)\times(\cN-1)$ identity matrix, and $\ket{u}$, $\ket{v}$ are $\cN-1$ dimensional vector valued fields. We choose  $\rho=(\cN-1)^{-1}$ in order to ensure that $\bb{U}_1\in\mathfrak{sl}_{\cN}$. It follows that $\bb{U}(\lambda)\in\mathfrak{sl}_{\cN}[\lambda]$. 

We assume that the Lax operator $\mathcal{L}(\lambda)$ is invariant under the $\mathbb{Z}_2$ reduction group generated by the involution 
\begin{equation}
r:\mathcal{L}(\lambda)\mapsto -Q\mathcal{L}(\lambda^*)^{\dagger}Q
\label{eq:redgr}
\end{equation}
where 
\begin{equation}
Q=\left(\begin{array}{cc}
\mathds{1} & 0 \\
0 & -\kappa 
\end{array}\right), \quad \kappa=\pm 1.
\end{equation}
The $\mathcal{L}(\lambda)^{\dagger}$ in \eqref{eq:redgr} denotes the formal adjoint operator of $\mathcal{L}(\lambda)$, i.e. $\mathcal{L}(\lambda)^{\dagger}=-D_x-\bb{U}(\lambda)^{\dagger}$. Moreover, $*$ denotes complex conjugation  and $\dagger$ stands for Hermitian conjugation. The invariance of $\mathcal{L}(\lambda)$ under $r$ implies that
\begin{equation}
-Q\bb{U}(\lambda^*)^{\dagger}Q=\bb{U}(\lambda).
\label{eq:Qsym}
\end{equation}
From equation \eqref{eq:Qsym} follows that
\begin{equation}
\alpha=\mathrm{i}a\in \mathrm{i}\mathbb{R}, \quad \beta=\sqrt{\kappa} \quad \text{and} \quad \ket{v}=\ket{u}^*\equiv \ket{u^*}.
\label{eq:reduction}
\end{equation}
For convenience we choose the normalisation $a=-(\r+1)^{-1}=(1-\cN)\cN^{-1}$.

It is easy to see that the operator 
\begin{equation}
\mathcal{A}(\lambda)=D_t-\bb{V}(\lambda), \quad \bb{V}(\lambda)=\lambda^2\bb{V}_2+\lambda \bb{V}_1+\bb{V}_0
\end{equation}
where
\begin{equation}
\bb{V}_2=-\bb{U}_1, \quad \bb{V}_1=-\bb{U}_0 \quad \text{and} \quad \bb{V}_0=\left(\begin{array}{cc}
\mathrm{i}\kappa\ketbra{u^*}{u} & -\mathrm{i}\sqrt{\k} \ket{u^*}_x \\
\mathrm{i} \sqrt{\k}\bra{u}_x & -\mathrm{i}\k\braket{u^*}{u}
\end{array}\right)
\label{eq:Vis}
\end{equation}
is also invariant under the action of $r$ and that the compatibility condition of the two operators
\begin{equation}
[\mathcal{L},\mathcal{A}]=0\Leftrightarrow \bb{U}_t-\bb{V}_x+[\bb{U},\bb{V}]=0
\label{eq:zcc}
\end{equation}
is equivalent to the vector NLS equation
\begin{equation}
\mathrm{i}\ket{u}_t+\ket{u}_{xx}-2\k|u|^2\ket{u}=0,
\label{eq:vNLS}
\end{equation}
where $|u|^2=\braket{u^*}{u}$.
Depending on which $\k$ we choose we obtain the focusing or defocusing vNLS equation (see \cite{Zakh-Shul, DegLomb1} for general focusing/defocusing systems of NLS equations). In what follows next we treat both vNLS equations simultaneously.

\subsection{Darboux transformation with a $\bb Z_2$ symmetry}

We are now in the position to construct Darboux-dressing transformations for the vNLS \eqref{eq:vNLS} using the reduction group \eqref{eq:redgr}. Let $\Psi(\l)$ be the fundamental solution of the linear system
\begin{equation}
\Psi_x=\bb{U}(\l)\Psi, \quad \Psi_t=\bb{V}(\l)\Psi
\label{eq:linearsyst}
\end{equation}
satisfying the initial condition $\Psi(\lambda)\big|_{(x,t)=(0,0)}=\mathds{1}$. Using the invariance condition \eqref{eq:Qsym} it follows that $Q\Psi(\l^*)^{\dagger^{-1}}Q$ is also a solution which satisfies the same initial condition, hence we obtain that $\Psi$ satisfies 
\begin{equation}
Q\Psi(\l^*)^{\dagger^{-1}}Q=\Psi(\l).
\label{eq:Psi-constraint}
\end{equation}

A Darboux transformation is a gauge transformation 
\begin{equation}
\Psi(\lambda)\mapsto\Phi(\lambda)=\bb{M}(\lambda)\Psi(\lambda).
\label{eq:DT}
\end{equation}
that leaves the linear system \eqref{eq:linearsyst} covariant. We call the matrix $\bb{M}(\lambda)$ Darboux or dressing matrix. We are interested to find those Darboux matrices so that $\Phi$ satisfies the same initial conditions as $\Psi$. This implies that $\Phi$ satisfies the constraint \eqref{eq:Psi-constraint} too. Then it is not hard to show that the Darboux matrix $\bb{M}(\lambda)$ has to satisfy the same relation \eqref{eq:Psi-constraint} or equivalently
\begin{equation}
\bb M(\lambda)^{-1}=Q\bb M(\lambda^*)^{\dagger}Q.
\label{eq:Minv1}
\end{equation}

The transformed fundamental solution $\Phi(\lambda)$ satisfies the linear system
\begin{equation}
\Phi_x=\widetilde{\bb{U}}(\l)\Phi, \quad \Phi_t=\widetilde{\bb{V}}(\l)\Phi
\label{eq:linearsyst1}
\end{equation}
where $\widetilde{\bb{U}}(\l)=\bb{U}({\ket{\ti u}},\l)$ and similarly for $\widetilde{\bb{V}}(\l)$. Using \eqref{eq:DT} and \eqref{eq:linearsyst}, equations \eqref{eq:linearsyst1} imply that the Darboux matrix satisfies the following equations 
\begin{equation}
\bb M_x=\widetilde{\bb{U}}\bb M-\bb M\bb{U}, \quad \bb M_t=\widetilde{\bb{V}}\bb M-\bb M\bb{V}
\label{eq:DLeq}
\end{equation}
known as Darboux-Lax equations. The Darboux-Lax equations are equivalent to the following relations
\begin{equation}
\widetilde{\mathcal{L}}(\lambda)=\bb M(\l)\mathcal{L}(\l)\bb M(\l)^{-1}, \quad \widetilde{\mathcal{A}}(\lambda)=\bb M(\l)\mathcal{A}(\l)\bb M(\l)^{-1},
\label{eq:dressing}
\end{equation}
where $\widetilde{\mathcal{L}}(\lambda)=D_x-\widetilde{\bb{U}}(\l)$ and $\widetilde{\mathcal{A}}(\lambda)=D_t-\widetilde{\bb{V}}(\l)$. Equations \eqref{eq:DLeq} are linear in $\bb M(\lambda)$ and thus invariant under a transformation of the form
$$
\bb M(\lambda)\mapsto f(\lambda)\bb M(\lambda)
$$
where $f(\lambda)$ is a non-zero scalar function of $\lambda$. This means that without any loss of generality we can assume that $\bb M(\lambda)$ has no poles at $\lambda=\infty$. The simplest such matrix with a single pole is of the form
\begin{equation}
\bb M(\l)=\mathds{1}+\frac{\bb M_0}{\l-\m}.
\label{eq:eldar}
\end{equation}

Equation \eqref{eq:Minv1} implies that the inverse of $\bb M(\l)$ has the form
\begin{equation}
\bb M(\lambda)^{-1}=\mathds{1}+\frac{Q\bb M_0^{\dagger}Q}{\lambda-\mu^*}.
\label{eq:Minv}
\end{equation}
Taking the residues at $\lambda=\mu$ and $\mu^*$ of equation $\bb M(\lambda)\bb M(\lambda)^{-1}=\mathds{1}$
we obtain
\begin{equation}
\bb M_0\left(\mathds{1}+\frac{Q\bb M_0^{\dagger}Q}{\mu-\mu^*}\right)=0, \quad \left(\mathds{1}+\frac{\bb M_0}{\mu^*-\mu}\right)Q\bb M_0^{\dagger}Q=0
\label{eq:eqfrompoles}
\end{equation}
respectively. Assuming that $\det (\bb M_0)\neq 0$ we have that $\det (Q\bb M_0^{\dagger}Q)\neq 0$ and thus the second equation of \eqref{eq:eqfrompoles} implies that $\bb M_0=(\mu-\mu^*)\mathds{1}$. In this case $\bb M(\lambda)=\frac{\l-\m^*}{\l-\m}\mathds{1}$ and so $\bb M$ is a trivial Darboux matrix. Hence we assume that $\bb M_0$ is not of full rank and specifically we are interested in the case where $\rank(\bb M_0)=1$  and thus $\bb M(\lambda)$ will be the simplest Darboux matrix (elementary Darboux matrix). In the case where the Lax matrices $\bb{U}$ and $\bb{V}$ are $2\times 2$ matrices the rank one case is the only possibility, however in our case $\rank(\bb M_0)=1,\ldots,\cN-1$. Nevertheless, we continue our investigations assuming that $\bb M_0$ has $\rank(\bb M_0)=s$.

Since $\rank(\bb M_0)=s$, $\bb M_0$ can be parametrised by two 
matrices of dimension $\cN\times s$
as
\begin{equation}
\bb{M}_0={\bf p}{\bf q}^T.
\label{eq:bi-vector}
\end{equation}  
Here $\textbf{q}=(q^1,\ldots,q^s)\in \text{M}_{\cN,s}(\bb C)$ with $q^j=(q^j_1,\ldots,q^j_{\cN})^T$ being $\cN$-vectors and similar for $\textbf{p}$.
Then we can solve \eqref{eq:eqfrompoles} with respect to $\bf{p}$ obtaining  
\begin{equation}
{\bf p}=(\mu-\mu^*)Q{\bf q}^*({\bf q}^TQ{\bf q}^*)^{-1}~.
\label{eq:ketp}
\end{equation}
Therefore, the Darboux matrix has the form
\begin{equation}
\bb M(\lambda)=\mathds{1}+\frac{\m-\m^*}{\l-\m}P,\quad \m\neq\m^*, \quad P=Q{\bf q}^*({\bf q}^TQ{\bf q}^*)^{-1}{\bf q}^T.
\label{eq:DM-P}
\end{equation}
Notice that $P$ is a projector $(P^2=P)$ and that the following relations 
\begin{equation}
\bb M(\m^*)Q{\bf q}^*=0, \quad {\bf q}^T\bb M(\m^*)=0, \quad \bb M^{-1}(\m)Q{\bf q}^*=0, \quad {\bf q}^T\bb M^{-1}(\m)=0~
\label{eq:properties}
\end{equation} 
are satisfied due to the following identities 
\begin{equation}
(\mathds{1}-P)Q\textbf{q}^*=0 \quad \text{and} \quad \textbf{q}^T(\mathds{1}-P)=0~.
\label{eq:properties0}
\end{equation}
Moreover, $P$ and thus $\bb M(\lambda)$ is invariant under the transformation 
\begin{equation}
{\bf q}\mapsto ~{\bf q}C, \quad C:(x,t)\mapsto C(x,t)\in GL(s,\mathbb{C}). 
\label{eq:rescaling}
\end{equation}
It follows that $\bb M(\lambda)$ is parametrised by a non-real complex number  $\mu$ and a point ${\bf q}$ in the complex Grassmannian  $Gr(s,\cN)\simeq \text{M}_{\cN,s}(\bb C)/GL(s,\bb C)$. In the next section we focus on the special case where $s=1$ and thus $\textbf{q}$ is an element of the projective space $\bb{P}^{\cN-1}(\bb{C})\simeq Gr(1,\cN)$. We note here that the $s=1$ case has been extensively used, see \cite{FT, Matveev, Cher, Nov} and references therein.

\subsection{Dressing and B\"acklund transformations}

We have used the symmetries of the Lax pair in order to write the general form of an elementary Darboux matrix \eqref{eq:DM-P}. Moreover the Darboux matrix must preserve the form of the Lax operators $\mathcal{L}$ and $\mathcal{A}$, i.e. equations \eqref{eq:dressing} must hold. Since the first part of \eqref{eq:dressing} must hold identically in $\l$ we obtain the following equations from the regular part at $\l=\infty$ and the residues at the simple poles at $\l=\m$ and $\l=\m^*$
\begin{equation}
\widetilde{\bb{U}}_0=\bb{U}_0+(\mu-\mu^*)\left[P,\bb{U}_1\right], \quad P\mathcal{L}(\mu)(1-P)=0, \quad (1-P)\mathcal{L}(\mu^*)P=0.
\label{eq:dres-x}
\end{equation}  
Using the fact that $P$ is a projector and of the form \eqref{eq:DM-P}, it is not hard to see that the second and the third equation of \eqref{eq:dres-x} are equivalent to the following eigenvalue problems 
\begin{equation}
{\bf q}^T_x+{\bf q}^T \bb{U}(\mu)={\bf f}~ {\bf q}^T, \quad {\bf q}^*_x - Q \bb{U}(\m^*) Q {\bf q}^* =  {\bf q}^* \, \textbf{f}_1 \,,
\label{eq:dres-x2}
\end{equation}
respectively, with $\textbf{f}$ and $\textbf{f}_1$  $s \times s$ matrix valued functions of $x$ and $t$. Taking into account the invariance of $\bb{U}$ under the reduction group \eqref{eq:Qsym} it follows that equations \eqref{eq:dres-x2} are compatible if $\textbf{f}_1=\textbf{f}^{\dagger}$.

Similarly, from the poles at $\l=\m$ and $\l=\m^*$ of the second equation of \eqref{eq:dressing} we obtain
$$
P\mathcal{A}(\mu)(1-P)=0 \quad \text{and} \quad (1-P)\mathcal{A}(\mu^*)P=0
$$
which imply that ${\bf q}$ also satisfy
\begin{equation}
{\bf q}^T_t + {\bf q}^T \bb{V}(\mu)= \textbf{g} \,{\bf q}^T , \quad {\bf q}^*_t - Q\bb{V}(\m^*) Q {\bf q}^* =  {\bf q}^*\, \textbf{g}_1 \,,
\label{eq:dres-t2}
\end{equation}
with $\textbf{g}, \textbf{g}_1$ being $s\times s$ matrix valued functions of $x,t$. Using again the invariance of $\bb{V}$ under the reduction group implies that $\textbf{g}_1=\textbf{g}^{\dagger}$. We have proved that ${\bf q}$ has to satisfy the system of equations
\begin{equation}
{\bf q}^T_x + {\bf q}^T \bb{U}(\mu) = \textbf{f} \, {\bf q}^T , \quad {\bf q}^T_t + {\bf q}^T \bb{V}(\mu) = \textbf{g} \, {\bf q}^T.
\label{eq:2eigenprobs}
\end{equation}
The compatibility of equations \eqref{eq:2eigenprobs} implies that $\textbf{f}$ and $\textbf{g}$ have to satisfy the zero curvature condition $\textbf{f}_t-\textbf{g}_x+\left[\textbf{f},\textbf{g}\right]=0$, and hence locally exists a matrix valued function $\textbf{h}$ such that $\textbf{f}=\textbf{h}_x\textbf{h}^{-1}$ and $\textbf{g}=\textbf{h}_t\textbf{h}^{-1}$. Since ${\bf q}\in Gr(s,\cN)$, the transformation
\begin{equation}
\textbf{q}\mapsto \textbf{q}\textbf{h}^T
\label{eq:fixing-base}
\end{equation}
preserves the form of the Darboux matrix and also makes equations \eqref{eq:2eigenprobs} homogeneous. Therefore, we obtain
\begin{equation}
{\bf q}^T = {\bf C}^T \Psi(\mu)^{-1} = {\bf C}^T Q \Psi(\mu^*)^{\dagger}Q
\label{eq:qfinal}
\end{equation}
where $\Psi(\mu^*)$ is the fundamental solution of the linear problem
\begin{equation}
\Psi_x=\bb{U}(\m^*)\Psi, \quad \Psi_t=\bb{V}(\m^*)\Psi
\label{eq:linprobm*}
\end{equation}
and ${\bf C}$ is a constant matrix of dimension $\cN\times s$.

Using \eqref{eq:qfinal} and the expression for $P$ \eqref{eq:DM-P},
we can write the projector matrix $P$ in terms of solutions of the linear system \eqref{eq:linprobm*} that correspond to the vNLS potential $\ket{u}=(u_1,\ldots,u_{\cN-1})^T$.
The first equation of \eqref{eq:dres-x} defines the transformation for the vNLS equation
\begin{equation}
\widetilde{u}_j=u_j+\frac{\mathrm{i}(\m^*-\m)}{\sqrt{\k}}P_{\cN j}, \quad j=1,\ldots,\cN-1
\label{eq:dressing-transf}
\end{equation}
and $P_{\cN j}$ can be written as a ratio of two determinants
\begin{equation}
P_{\cN j}=\frac{
\left|
\begin{array}{cccc}
0 & \k q_{\cN}^{1^*} & \cdots & \k q_{\cN}^{s^*}    \\
q_{j}^1 & q^{1^T}Qq^{1^*} & \cdots & q^{1^T}Qq^{s^*} \\
\vdots  &      \vdots     & \ddots &      \vdots      \\
q_{j}^s & q^{s^T}Qq^{1^*} & \cdots &    q^{s^T}Qq^{s^*}
\end{array}
\right|
}{\left|
\begin{array}{ccc}
q^{1^T}Qq^{1^*} & \cdots & q^{1^T}Qq^{s^*} \\
   \vdots     & \ddots &      \vdots      \\
 q^{s^T}Qq^{1^*} & \cdots &    q^{s^T}Qq^{s^*}
\end{array}
\right|}\, , 
\label{eq:ratiodets}
\end{equation}
where the $q^{i}$ $i=1,\ldots s$ denote the columns of matrix $\textbf{q}$ \eqref{eq:qfinal}. Specific solutions for various different $s$ will be presented elsewhere. 

The special case $s=\rank{\bb M_0}=1$ is the simplest and has additional interest. In this case
$P$ takes the form
\begin{equation}
P=\frac{Q\textbf{q}^*\textbf{q}^T}{\textbf{q}^TQ\textbf{q}^*}
\label{eq:Ps=1}
\end{equation}
where $\textbf{q}$ is now an $\cN$-vector. Moreover, equation \eqref{eq:dressing-transf}  takes the form
\begin{equation}
\widetilde{u}_j=u_j-\rm i (\m^*-\m)\sqrt{\k} \frac{q^*_{\cN} q_j}{|q_1|^2 + \cdots+|q_{\cN - 1}|^2- \k|q_{\cN}|^2}, \quad j=1,\ldots, \cN-1
\label{eq:vNLSdress}
\end{equation}
which is the dressing transformation for both the focusing and defocusing vector NLS equation  \cite{Park-Shin, Wright}.  The transformation \eqref{eq:vNLSdress} is a generalisation of the well known dressing transformation for the scalar NLS equation, see \cite{Z-S, Its-Salle, FT}. 

In the rank$\ \mathbb{M}_0 =1$ case we can also use the Darboux matrix \eqref{eq:DM-P} in order to derive the B\"acklund transformations for both focusing and defocusing vNLS equations and for arbitrary $\cN$. To this end we first use the rescaling symmetry \eqref{eq:rescaling} and write ${\bf q}$ in the following form
\begin{equation}
{\bf q}=\left( 
\begin{array}{c}
\ket{q} \\
1
\end{array}
\right).
\label{eq:rescalledq}
\end{equation}
Then $\bb M_0$ takes the form
\begin{equation}
\bb M_0=\left(
\begin{array}{cc}
d \ketbra{q^*}{q} & d \ket{q^*} \\
-\k d\bra{q} & -\k d
\end{array}
\right)
\label{eq:rescalledM0}
\end{equation}
where
\begin{equation}
d=\frac{\m-\m^*}{\abs{q}^2-\k} \quad \text{with} \quad d^*=-d~.
\label{eq:d-factor}
\end{equation}

The first equation of \eqref{eq:dres-x} implies that
\begin{equation}
\ket{q}=\frac{1}{\mathrm{i}\sqrt{\k}d}(\ket{\ti u}-\ket{u}).
\label{eq:back-q}
\end{equation}
It follows that
\begin{equation}
\abs{q}^2=\braket{q^*}{q}=-\frac{1}{d^2}\abs{\ti u-u}^2
\label{eq:modq}
\end{equation}
and using \eqref{eq:d-factor} we obtain that $d\neq 0$ satisfies the quadratic equation
\begin{equation}
\k d^2-(\m^*-\m)d+\abs{\ti u-u}^2=0~.
\label{eq:quadratic}
\end{equation}
Therefore, 
\begin{equation}
d=\frac{\m^*-\m}{2\k}\pm\eta
\label{eq:d+-}
\end{equation}
where
\begin{equation}
\eta=\sqrt{\left(\frac{\m^*-\m}{2}\right)^2-\k\abs{\ti u-u}^2}.
\label{eq:eta}
\end{equation}

From the pole at $\l=\m$ of equations \eqref{eq:DLeq} we see that $\bb M_0$ has to satisfy
\begin{equation}
\bb M_{0x}=\ti{\bb{U}}(\mu)\bb M_0-\bb M_0\bb{U}(\mu) \quad \text{and} \quad \bb M_{0t}=\ti{\bb{V}}(\mu)\bb M_0-\bb M_0\bb{V}(\mu)~.
\label{eq:diffeqM0}
\end{equation}
The first equation of \eqref{eq:diffeqM0} implies 
\begin{equation}
(-\k d\bra{q})_x=-\mathrm{i}d \m \k \bra{q}+\sqrt{\k}d\braket{\ti u}{q^*}\bra{q}+\k\sqrt{\k}d\bra{u}
\label{eq:Back-x-1}
\end{equation}
while from the second we obtain
\begin{equation}
(-\k d \bra{q})_t = \m (\k d\bra{q})_x+\mathrm{i}\sqrt{\k}d\braket{\ti u_x}{q^*}\bra{q}+\mathrm{i}d\abs{\ti u}^2\bra{q}+\mathrm{i}d\braket{q}{u^*}\bra{u}+\mathrm{i}\k\sqrt{\k}d\bra{u_x}~.
\label{eq:Back-t-1}
\end{equation} 
Using relation \eqref{eq:back-q} and \eqref{eq:d+-} equation \eqref{eq:Back-x-1} takes the form
\begin{equation}
\mathrm{i}\left(\ket{\ti u}-\ket{u}\right)_x=-\m\left(\ket{\ti u}-\ket{u}\right)+\left(\frac{\m^*-\m}{2}\pm\eta\right)\ket{u}-\frac{\abs{\ti u}^2-\braket{\ti u}{u^*}}{\abs{\ti u-u}^2}\left(\frac{\m^*-\m}{2}\mp\eta\right)\left(\ket{\ti u}-\ket{u}\right)
\label{eq:Back-x-2}
\end{equation}
and constitutes the $x-$part of the B\"acklund transformation of vNLS while \eqref{eq:Back-t-1} can be rewritten as 
\begin{align}
\mathrm{i}(\ket{\ti u}-\ket{u})_t &= - \mathrm{i} \m (\ket{\ti u}-\ket{u})_x-\mathrm{i}\left(\frac{\braket{\ti u_x}{\ti u^*}-\braket{\ti u_x}{u^*}}{\abs{\ti u-u}^2}\right)\left(\frac{\m^*-\m}{2}\mp\eta\right)(\ket{\ti u}-\ket{u})  \nonumber \\
& +  \k\abs{\ti u}^2(\ket{\ti u}-\ket{u}) + \k \left(\braket{\ti u}{u^*}-\abs{u}^2\right)\ket{u}+\mathrm{i} \left( \frac{\m^* - \m}{2} \pm \eta \right) \ket{u_x}~.
\label{eq:Back-t-2}
\end{align}
When $\cN = 2$ the B\"acklund transformation \eqref{eq:Back-x-2}, \eqref{eq:Back-t-2} becomes the known BT for NLS equation with $\k = \pm 1$ (see \cite{boitipemp}).

\subsubsection*{B\"acklund transformations: solitons to anti-solitons}
We shall briefly discuss here the existence of a novel type of B\"acklund transformations that associate solitonic to anti-solitonic solutions.
This idea is essentially inspired by  the existence of certain boundary conditions in high rank $gl_{\cal N}$ integrable systems that force a soliton to reflect as an anti-soliton. In the language of representation theory in quantum integrable systems this translates into a change of the representation of a particle into its conjugate after reflection. In the present context and associated to the notion of ``integrable defects'', that we are interested in, such a BT may be seen as a discontinuity in the one dimensional system relating solitonic to anti-solitonic solutions of the non-linear equation at hand. In a more physical context the defect as a quasi BT can be thought of as a transmitting object that turns each field to its conjugate after transmission. This is mathematically and physically a very  interesting issue, which will be discussed in detailed in future works. Nevertheless,  we shall give a first flavour of this behaviour here.

Let us introduce the following object:
\be
\ti{\bb{U}}(X,\l) = - \bb U^T(X,-\l),
\ee
where $^T$ denotes usual transposition.
Then the corresponding time component of the Lax pair can be derived via the familiar formula below \cite{sts}
\be
\tilde {\bb V}(x,\l,\m) = \tau^{-1}(\l)\, \tr_a \Big( \tilde {\bb T}_a(- L, x, \l) \, r^{T_a T_b}_{ab} (-\l + \m) \, \tilde {\bb T}_a (x, L, \l) \Big) \,,
\ee
where $^{T_a}$ denotes transposition on the space characterized by the index $a$, and
\ba
&&\tau(\l)=tr \tilde {\bb T}(-L, L;\l),~~~~\tilde {\bb T}(a,b,\l) = {\bb T}^T(b,a,\l),~~~b > a \cr
&&{\bb T}(b,a,\l)= \exp\Big \{ \int_a^b dx\ {\mathbb U}(x,\l)\Big \}.
\ea 
In the case where $r$ is the Yangian matrix the $r_{ab}^{T_a T_b} = r_{ab}$. Working out the BT for the setting above we end up to structurally similar BTs as the ones defined earlier in the text, but now the following identifications hold:
\be
\l \to -\l, ~~~~|\tilde u \rangle  \to |u^*\rangle, ~~~~|\tilde u^* \rangle  \to |u\rangle.
\ee
In the vNLS case the situation is quite straightforward, however more interesting and presumably richer scenarios could arise in more involved models, such as the affine Toda field theories or higher rank Landau-Lifshitz models. Also, this setting naturally applies to discrete integrable modes associated to higher rank algebras. All these are significant issues that will be discussed in detail in future investigations, given that our main purpose here is to provide a brief introduction to the soliton anti-soliton type BTs.

\subsection{Higher Darboux transformation}

In this section we investigate Darboux-dressing transformations that correspond to multi-soliton solutions. In principle, in order to obtain higher soliton solutions one can consider compositions of elementary Darboux transformations of the form \eqref{eq:eldar} with several different poles in $\lambda$, see \cite{FT, Nov}. However, here we are interested in a non-elementary Darboux matrix, which has $n$ poles and is of the form
\begin{equation}
\bb M(\lambda)=\mathds{1}+\sum_{i=1}^n\frac{\bb M_i}{\l-\m_i} \, .
\label{eq:DarMat-n}
\end{equation} 
Moreover, we assume that $\bb M(\lambda)$ has the same structure as the 1-soliton Darboux matrix, that is it satisfies relation \eqref{eq:Minv1}. It follows that the inverse matrix is of the form
\begin{equation}
\bb M(\lambda)^{-1}=\mathds{1}+\sum_{i=1}^n\frac{Q\bb M_i^{\dagger}Q}{\l-\m_i^*}.
\label{eq:Minv2}
\end{equation}
Comparing the asymptotic expansions of $\bb M(\l)$ and $\bb M(\l)^{-1}$ at $\l \rightarrow \infty$ we also obtain the following relation
\be \label{sum Mi}
\sum_{i=1}^{n} \bb M_i = -  \sum_{i=1}^{n} Q \bb M_i^{\dagger} Q \,.
\ee 
Taking the residue at $\lambda=\m_j$ and $\m_j^*$ of equation $\bb M(\l)\bb M(\l)^{-1}=\mathds{1}$ we have that
\begin{equation}
\bb M_j\bb M(\m_j)^{-1}=0,\quad \bb M(\m_j^*)Q\bb M_j^{\dagger}Q=0, \quad j=1,\ldots n,
\label{eq:resmj}
\end{equation}
respectively.
The above equations imply that all $\bb M_j$ are not of full rank. In general we can proceed assuming that $\rank(\bb M_j)=s_j$ with $1 \le s_j\le \cN-1$ but instead we will treat only the case where $\rank(\bb M_j)=1$ for all $j$.

As in the 1-soliton case, we can express all $\bb M_j$ in the form $\bb M_j=\textbf{p}_j\textbf{q}_j^T$ where $\textbf{p}_j$ and $\textbf{q}_j$ are $\cN$-vectors. Then, equations \eqref{eq:resmj} imply that
\begin{equation}
\textbf{q}_j^T\bb M(\m_j)^{-1}=0, \quad \bb M(\m_j^*)Q\textbf{q}_j^*=0, \quad j=1,\ldots,n
\label{eq:resmj-2}
\end{equation}
respectively. As the relations in \eqref{eq:resmj-2} are equivalent to each other, using one of them we have that
\begin{equation}
\sum_{i=1}^n\frac{\textbf{q}_i^T Q \textbf{q}_j^*}{\m_i-\m_j^*} \textbf{p}_i  =  Q \textbf{q}_j^*, \quad j=1,\ldots,n\, 
\label{eq:kereq-q}
\end{equation}
with $ \m_i \neq \m_j^*$ for all $i,j$. We define the scalar quantities $(\textbf{q}_i,\textbf{q}_j)=\frac{\textbf{q}_i^TQ\textbf{q}_j^*}{\m_i - \m_j^*}$ and if we assume that the Cauchy type matrix $(\textbf{q}_i,\textbf{q}_j)$ is invertible, then  using the Cramer's rule we can solve \eqref{eq:kereq-q} for all $\textbf{p}_i$. In  this way the $\textbf{p}_i$'s can be expressed in terms of the ${\bf q}_i$'s as a ratio of determinants
\begin{equation}
\textbf{p}_i=\frac{\left|\begin{array}{ccccccc}
(\textbf{q}_1,\textbf{q}_1) & \cdots & (\textbf{q}_1,\textbf{q}_{i-1}) & Q \textbf{q}_1^* & (\textbf{q}_1,\textbf{q}_{i+1}) & \cdots & (\textbf{q}_1,\textbf{q}_{n})\\
\vdots &\ddots & \vdots & \vdots & \vdots & \ddots & \vdots\\
(\textbf{q}_n,\textbf{q}_1) & \cdots & (\textbf{q}_n,\textbf{q}_{i-1}) &  Q \textbf{q}_n^* & (\textbf{q}_n,\textbf{q}_{i+1}) & \cdots & (\textbf{q}_n,\textbf{q}_{n})
\end{array}\right|} {\left|\begin{array}{ccc}
(\textbf{q}_1,\textbf{q}_1) & \cdots & (\textbf{q}_1,\textbf{q}_n) \\
\vdots & \ddots & \vdots \\
 (\textbf{q}_n,\textbf{q}_1) & \cdots & (\textbf{q}_n,\textbf{q}_n)
\end{array}
\right|}.
\label{eq:sol-linear}
\end{equation}
The symbolic determinant in the numerator is expanded with respect to the i-th column.

From the regular part at $\l = \infty$ of the dressing relations \eqref{eq:dressing} we obtain
\be  \label{dressing higher}
\ti {\bb U}_0 =  \bb U_0 + \comm{\sum_{i =1}^{n} {\bf p}_i {\bf q}_i^{T} }{ \bb U_1} \,, 
\ee
where we have used relation \eqref{sum Mi}, while from the residue at $\l = \m_j^*$ we have that
\be \label{eq:diff res m}
 \bb M(\m_j^*) \mathcal{L}(\m_j^*) Q \bb M_j^{\dagger} Q = 0 \,, \quad \bb M(\m_j^*) \mathcal{A}(\m_j^*) Q \bb M_j^{\dagger} Q = 0 \,.
\ee
Similar to the single pole case, using relations \eqref{eq:resmj-2}, equations \eqref{eq:diff res m} imply that 
\be
{\bf q}_{j x} - Q \bb U (\m_j^*)^* Q {\bf q}_j = 0 \,, \quad {\bf q}_{j t} - Q \bb V (\m_j^*)^* Q {\bf q}_j = 0 \,,
\ee
hence we can write
\be
{\bf q}_j^{T} = C_j^{T} Q \Psi (\m_j^*) ^{\dagger} Q \,,
\ee
where the $C_j$ are constant $\cN$-vectors and $\Psi (\m_j^*) $ is a fundamental solution to the linear problem at $\l = \m_j^*$.

Using equation \eqref{eq:sol-linear}, relation \eqref{dressing higher} is the dressing transformation which can be written as
\be \label{dressing comp higher}
\ti{u}_i = u_i - \frac{\text{i}}{\sqrt{\k}} \frac{\tau_i}{\tau} \,, \quad i = 1, \cdots, \cN - 1 \,,
\ee
where $\tau, \tau_i$ stand for the following determinants
\be
\tau = \left|\begin{array}{ccc}
(\textbf{q}_1,\textbf{q}_1) & \cdots & (\textbf{q}_1,\textbf{q}_n) \\
\vdots & \ddots & \vdots \\
 (\textbf{q}_n,\textbf{q}_1) & \cdots & (\textbf{q}_n,\textbf{q}_n)
\end{array}
\right|  ,
\quad 
\tau_i =  \left|\begin{array}{cccc}
0 & {\bf q}_{1,i} & \cdots  & {\bf q}_{n,i}  \\
\k {\bf q}_{1, \cN}^* & (\textbf{q}_1,\textbf{q}_1) & \cdots & (\textbf{q}_1,\textbf{q}_n) \\
\vdots & \vdots & \ddots & \vdots \\
\k {\bf q}_{n, \cN}^* & (\textbf{q}_n,\textbf{q}_1) & \cdots & (\textbf{q}_n,\textbf{q}_n) 
\end{array}
\right|  ,
\ee
with ${\bf q}_{k,m}$ denoting the $m$-th component of the ${\bf q}_k$ vector. Recently, bright and dark soliton solutions were obtained using the dressing method, see \cite{Tsu, china}.

\section{Integral operators as global Darboux transformations $\&$ dressing}

We shall focus on situations where the dressing is expressed in terms of integral representations. Let us recall the ``bare'' differential operators associated to the vector NLS model (see e.g. \cite{Z-S nls, Drazin} and references therein):
\ba
&&{\cal D}_0^{(1)} = {\cal M} \ {\partial}_x,
~~~~ {\cal M} = \sum_i \alpha_i\ e^{({\cal N})}_{ii} \cr
&& {\cal D}_0^{(2)}= (i a\
\partial_t-\partial^2_x)\mathds{1}\label{D0}
\ea
where recall $e_{ij}^{({\cal N})}$ are ${\cal N} \times {\cal N}$ matrices with elements $(e_{ij}^{({\cal N})})_{kl} = \delta_{ik}\delta_{jl}$.

Let us briefly recall the Zakharov-Shabat (ZS) dressing \cite{Z-S, Drazin}, which is equivalent to the inverse scattering transform as well as the Riemann-Hilbert problem, and leads to the Gelfand-Levitan-Marchenko (GLM) equation (see \cite{FT, Drazin, booknls}):
\be
K(x, z) + F(x,z) + \int_{x}^{\infty} K(x,y) F(y,z) dz =0,~~~~~x < z. \label{glm}
\ee
The starting point of the formulation is the following factorization for the operator $\mathds{1} +{\cal F}$, which holds at the abstract operator level:
\be
(\mathds{1}+ {\cal K})\ (\mathds{1} + {\cal F}) = (\mathds{1}+ \hat {\cal K}),
\ee
and we define the integral representations as:
\ba
&&{\cal F}(\mathrm{f}) = \int_{-\infty}^{\infty} F(x,y) \mathrm{f}(y)dy \cr
&& {\cal K}(\mathrm{f}) = \int_{x}^{\infty} K(x, y) \mathrm{f}(y) dy \cr
&& \hat {\cal K}(\mathrm{f}) = \int_{-\infty}^x \hat K(x, y) \mathrm{f}(y) dy. \label{integral}
\ea
In the matrix language the factorization of $I +{\cal F}$ corresponds essentially to the decomposition on upper and lower triangular matrices. Note also that $F$ satisfies the linear equations emanating from the following invariant actions:
\be
{\cal D}_0^{(i)}\ {\cal F} = {\cal F}\ {\cal D}_0^{(i)}, ~~~~i \in \{1,\ 2\}.
\ee
Thus via the integral representation of ${\cal F}$ the following equations arise
\ba
&& i a\ F_t -F_{xx}+ F_{zz} =0 \cr
&& M\ F_x +  F_z\ M =0
\ea
where
\be
F(x,z;t) = \sum_{j=2}^{\cal N} f_j(x,z;t)\ e^{({\cal N})}_{1j}+ \sum_{j=2}^{\cal N} \hat f_j(x,z;t)\ e^{({\cal N})}_{j1}. \label{FF}
\ee
It is worth noting that a more general choice of the solutions of the linear problem i.e $F$ expressed in the generic Grassmannian form:
\ba
F = \begin{pmatrix} {\bf 0}_{k \times k} & {\mathbb X}_{k\times{\cal N}} \\
{\mathbb Y}_{{\cal N}\times k }& {\bf 0}_{{\cal N}\times {\cal N}} \end{pmatrix}
\ea 
will provide solutions to the matrix NLS model, however this problem will be discussed in detail elsewhere.

We shall henceforth focus on solutions of the linear equations above (\ref{FF}) that are factorizable i.e.,
\ba
&&f_j(x,z;t) = \sum_{\alpha=1}^n X_j^{(\alpha)}(x,t) Z_j^{(\alpha)}(z)\cr
&&\hat f_j(x,z;t) = \sum_{\alpha=1}^n  \hat X_j^{(\alpha)}(x,t) \hat Z_j^{(\alpha)}(z).\
\ea
It is clear that the ZS dressing may be thought of as a global Darboux transformation; it is essentially a Darboux transformation in an integral representation:
\ba
&& \Psi = {\cal B}\ \Psi_0, \cr
&& \Psi =  \mathds{1} + \hat {\cal K},
~~~~\Psi_0 = \mathds{1} + {\cal F},
~~~~{\cal B}= \mathds{1} + {\cal K}.
\ea

We come now to the main objective, which is the solution of the GLM equation (\ref{glm}) for the vector NLS system. Given the form of the solution $F$ (\ref{FF}), and also considering the generic expression $K(x,z) = \sum_{i,j} K_{ij}(x,z) e_{ij}$  we end up to the following set of equations (see also \cite{ablowitz, Drazin} for the $sl_2$ NLS case):
\ba
&& K_{1j}(x,z) + f_j(x,z) + \int_{x}^{\infty} K_{11}(x,y)f_j(y,z)\ dy=0\cr
&& K_{11}(x,z) + \int_x^{\infty}\sum_{j} K_{1j}(x,y )\hat f_j(y,z)\ dy=0, \label{basic1}
\ea
\ba
&& K_{i1}(x,z) + \hat f_i(x,z) + \sum_j\int_{x}^{\infty} K_{ij}(x,y,t)\hat f_j(y,z)\ dy = 0\cr
&& K_{ij}(x,z) + \int_x^{\infty} K_{i1}(x,y,t) f_j(y,z)\ dy = 0. ~~~j\in\{2, \ldots, {\cal N}\}. \label{basic2}
\ea
The two sets of equations above can be independently solved to provide $K_{1j},\ K_{11}$ and $K_{i1},\ K_{ij}$ respectively. Moreover, given the form of the dressed operators it is clear that $K_{1j}$ and $K_{j1}$ provide the fields $u_{j-1}$ and $u^*_{j-1}$ respectively, i.e the components of $\langle u |$ and $|u^* \rangle$ (see also \cite{zen-elim}).

Solving the latter system (\ref{basic1}) we obtain
\be
K_{1j}(x,z) + f_j(x,z) -\int_{x}^{\infty} \int_x^{\infty}dy\ d \tilde y \sum_i K_{1i}(x, \tilde y) \hat f_i(\tilde y, y) f_j(y,z)=0.
\ee
Due to the form of the latter formula we can consider the following factorization of the kernel $K_{1j}$
\be
K_{1j}(x,z;t)= \sum_{\alpha=1}^n L_{j}^{(\alpha)}Z_j^{(\alpha)}(z).
\ee
Recalling the form of $f_j,\ \hat f_j$ and after integration we end up with the following generic expression:
\be
\sum_{i} \sum_{\beta}L_i^{(\beta)}(x,t) {\mathbb M}_{ij}^{\beta \alpha }  = -X_j^{(\alpha)}(x,t) \label{linear1}
\ee
where we define:
\be
{\mathbb M}_{ij}^{\beta \alpha } = \delta_{ij} \delta_{\alpha \beta} -P_{ii}^{\beta \gamma}\ \hat P_{ij}^{\gamma \alpha}
\ee
and
\be
P_{ii}^{\beta \gamma}(x,t) = \int_{x}^{\infty}dy\ Z_i^{(\beta)}(y)\hat X_{i}^{(\gamma)}(y, t),~~~~
\hat P_{ij}^{\gamma \alpha}(x,t) = \int_{x}^{\infty}dy\ \hat Z_i^{(\gamma)}(y) X_{j}^{(\alpha)}(y,t). \label{PP}
\ee
Notice that the obvious choice
\ba
&& X_j^{(\alpha)}(x,t) = b_j e^{i\Lambda_j^{(\alpha)}t+i\lambda_j^{(\alpha)}x}, ~~~~Z_j^{(\alpha)}= e ^{i \mu_j^{(\alpha)}z}\cr
&& \hat X_j^{(\alpha)}(x,t) =\hat b_j e^{i\hat \Lambda_j^{(\alpha)}t+i\hat \lambda_j^{(\alpha)}x}, ~~~~\hat Z_j^{(\alpha)} = e ^{i \hat \mu_j^{(\alpha)}z}
\ea
leads to simple expressions for $P,\ \hat P$ after integration (see also \cite{booknls} for relevant expressions in the context of the inverse scattering transform):
\ba
&& P_{jj}^{\beta \gamma} = - \hat b_j^{(\gamma)} {e^{i\hat \Lambda_j^{(\gamma)}t+i\hat \lambda_j^{(\gamma)}x + i \mu_j^{(\beta)}x} \over i(\hat \lambda_j^{(\gamma)} + \mu_j^{(\beta)})}\cr
&& \hat P_{ij}^{\gamma \alpha}=- b_j^{(\alpha)} {e^{i \Lambda_j^{(\alpha)}t+i \lambda_j^{(\alpha)}x + i\hat  \mu_i^{(\gamma)}x} \over i(\lambda_j^{(\alpha)} + \hat \mu_i^{(\gamma)})}. \label{pp2}
\ea
Equation (\ref{basic1}) can be expressed in a more compact form as:
\be
{\mathbb L} \cdot {\mathbb M} = -{\mathbb X} \Rightarrow {\mathbb L} = -{\mathbb X} \cdot {\mathbb M}^{-1}, \label{lm}
\ee
where
\be
{\mathbb L} = \sum_{\alpha} \sum_{j} L_{j}^{(\alpha)}\hat e_{j}^{*({\cal N})} \otimes \hat e_{\alpha}^{*(n)},~~~~~{\mathbb M} = \sum_{\alpha, \beta}\sum_{i, j}{\mathbb M}_{ij}^{\alpha \beta} e_{ij}^{({\cal N})} \otimes e_{\alpha \beta}^{(n)}, ~~~{\mathbb X} =\sum_{\alpha}\sum_{j} X_j^{(\alpha)} \hat e_{j}^{*({\cal N})} \otimes e_{\alpha}^{*(n)}
\ee
and $e_j^{*({\cal N})}$ are the ${\cal N}$ dimensional column vectors with 1 at position $j$ and zero elsewhere. Our task is to identify $K_{1j}$, which will provide in turn the fields $u_{j-1}$; indeed $K_{1j}(x,x) \propto u_{j-1}(x)$ \cite{booknls,zen-elim}.

Similarly, we solve the system (\ref{basic2}) to identify the quantities $K_{i1}$, which in turn provide the fields $u^{*}_{i-1}$. From the form of the system it is easier first to solve for $K_{ij}$ ($i, j \neq 1$) and then obtain $K_{i1}$, and hence the $ u^*_{i-1}$ fields (time dependence is implicit in all the expressions below)
\be
K_{ij}(x,z) - \int_{x}^{\infty} \hat f_i(x,y) f_j(y,z) dy
- \sum_l \int_x^{\infty} \int_{x}^{\infty}d\tilde y\ dy\
K_{il}(x,\tilde y)\hat f_l(\tilde y, y) f_j(y,z). \label{kk}
\ee
From the latter expression and the chosen form of the solution of the linear system we can consider the following factorized form for $K_{ij}$:
\be
K_{ij}(x,z) = \sum_{\alpha=1}^n L_{ij}^{(\alpha)}(x,t)\ Z^{(\alpha)}_j(z).
\ee
Substituting the latter expression in (\ref{kk}) we obtain the following linear system:
\be
\sum_{l} \sum_{\beta}L_{il}^{(\beta)} {\mathbb M}^{\beta \alpha}_{lj} = \sum_{\beta}\hat X_i^{(\beta)}(x) \hat P_{ij}^{\beta \alpha}(x). \label{lm2}
\ee
Moreover, the quantities $K_{11}$ and $K_{ij}$ may be also derived via (\ref{basic1}), (\ref{basic2}), hence we obtain:
\ba
&& K_{11}(x, z) = -\sum_{j}\sum_{\beta, \gamma} L_j^{(\beta)}(x)P_{jj}^{\beta \gamma}(x) \hat Z_j^{(\gamma)}(z) \cr
&& K_{ij}(x, z) = -\sum_{\beta}\hat X_i^{(\beta)}(x)\hat Z_i^{(\beta)}(z)-\sum_{j} \sum_{\beta, \gamma} L_j^{(\beta)}(x)P_{jj}^{\beta \gamma}(x) \hat Z_j^{(\gamma)}(z).
\ea

It is easy now to extract for instance the one-soliton solution given the description above. Indeed expressions (\ref{lm}),(\ref{lm2}) still hold, but now ${\mathbb M}$ is defined as (no Greek letter indices involved any more as is natural):
 \be
{\mathbb M}_{ij} = \delta_{ij}  -P_{ii}\ \hat P_{ij}
\ee
where we define:
\be
P_{ii}(x,t) = \int_{x}^{\infty}dy\ Z_i(y)\hat X_{i}(y, t),~~~~
\hat P_{ij}(x,t) = \int_{x}^{\infty}dy\ \hat Z_i(y) X_{j}(y,t). \label{PP2}
\ee
For the sake of simplicity let us consider the case where all the spectral  parameters $\lambda_j$ are the same for all the fields, then it is clear that:
\be
f_j(x,z;t) = b_j e^{i\Lambda t + i \lambda x + i\mu z},
~~~~\hat f = \hat b_j e^{i\Lambda t + i \hat \lambda x + i\hat \mu z}
\ee
$b_j,\ \hat b_j$ are the components of the so called {\it polarization vectors}.
Then it is clear from the integral equations $K_{1j},\ K_{j1}$:
\be
K_{1j}(x,z;t) = L_j(x,t)e^{i\mu z}, ~~~K_{j1}(x,z;t) =
\hat L_j(x,t) e^{i\hat \mu z}.
\ee
One can then easily obtain a solution for $\hat L_j$. Indeed, one obtains a simple scalar equation, which immediately provides the solution
\ba
&& \hat L_j(x,t)  = - {\hat b_j e^{i\hat \Lambda t + i \hat \lambda x} \over 1 + {\mathrm C}\ {\mathrm H}(x,t)}, \cr
&& \mbox{where} ~~~{\mathrm C} = \sum_{j} b_j\ \hat b_j, ~~~~{\mathrm H}(x,t) = {e^{i(\hat \Lambda +\Lambda) t + i (\lambda + \mu +\hat \lambda + \hat \mu) x}\over (\lambda +\hat \mu)(\hat \lambda+ \mu)}.
\ea
Similarly, the expressions for $L_j$ reduce into the simple formulas below:
\be
{\mathbb L} = -e^{i\Lambda t + i \lambda x}\ {\mathbb B}\ {\mathbb M}^{-1}
\ee
where we define:
\be
{\mathbb L} = \sum_{j=1}^N L_j\hat e_{j}^{*({\cal N})}, ~~~~~~{\mathbb B} =\sum_{j=1}^N b_j\hat e_{j}^{*({\cal N})}, ~~~~~{\mathbb M} = \mathds{1} + {\mathbb P}
\ee
and ${\mathbb P}$ is expressed as a bi-vector 
\be
{\mathbb P} = {\mathrm H}(x,t)\ \hat {\mathbb B}^T {\mathbb B}
\ee
and is also a projector:
\be
{\mathbb P}^{2} = {\mathrm C}\ {\mathrm H}(x,t)\ {\mathbb P},
\ee
which leads to the immediate identification of the inverse ${\mathbb M}^{-1}$
\be
{\mathbb M}^{-1} = \mathds{1} - {1 \over 1 +{\mathrm C}\ {\mathrm H}(x,t)} {\mathbb P}.
\ee
The identification of $\hat L_i$ is then straightforward
\be
L_j(x,t) = -{b_j e^{i\Lambda t + i \lambda x}\over 1 + {\mathrm C}\ {\mathrm H}(x,t)},
\ee
and clearly compatible with the solution for $\hat L_j(x,t)$.

\subsubsection*{The continuous case}

It will be instructive for the general purposes of studying solutions of integrable PDEs, but also in association with the time evolution of point-like defects to consider the continuum case. Basically the structure of the linear equations emanating from GLM remains intact, however instead of the matrix formulation one employs in this case linear integral equations as will be evident below.

The essential difference with the discrete case studied above is that all discrete sums formally turn into integrals i.e
\be
\sum_{\alpha =1}^n f^{(\alpha)} \to \int_{-\infty}^{\infty}dk\ f(k).
\ee
More precisely, considering factorized expressions for the solutions of the linear problem
\ba
&&f_j(x,z;t) = \int_{-\infty}^{\infty}dk\ X_j(k;x,t)\ Z_j(k; z) \cr
&& \hat f(x,z;t) = \int_{-\infty}^{\infty}dk\ \hat X_j(k;x,t)\ \hat Z_j(k; z)
\ea
we then obtain the continuum limit for the factorization of $K_{1j}$:
\be
K_{1j}(x,z,t)= \int_{-\infty}^{\infty}dk\ L_j(x,t;k)Z_j(z;k).
\ee
The fundamental linear equation (\ref{linear1}) is then written as
\be
\sum_i \int_{-\infty}^{\infty}d\tilde k\ L_i(x,t; \tilde k)\ {\mathbb M}_{ij}(x,t;\tilde k, k) = -X_j(x,t;k),
\ee
where we define
\be
{\mathbb M}_{ij}(\tilde k,\ k) = \delta_{ij} \delta(\tilde k, k ) -\int dk' P_{ii}(\tilde k,\ k')\ \hat  P_{ij}(k', k)
\ee
and $P,\ \hat P$ are then defined as the continuum analogues of (\ref{pp2}), i.e.
\ba
&& P_{ii}(x,t; k, \tilde k) = {\hat b_j(\tilde k)e^{i\hat \Lambda(\tilde k)t + i \hat \lambda(\tilde k)x + i \mu(k)x} \over i (\hat \lambda(\tilde k) + \mu(k)) } \cr
&& \hat P_{ij}(x,t; k, \tilde k) = {b_j(\tilde k)e^{i \Lambda(\tilde k)t + i \lambda(\tilde k)x + i  \hat\mu(k)x} \over i ( \lambda(\tilde k) +  \hat \mu(k)) }.
\ea

Similarly, as in the discrete case we can obtain the factorized form:
\be
K_{ij}(x,z) = \int dk\ L_{ij}(x,t; k) Z_j(z;k)
\ee
and the respective linear equation
\be
\sum_l\int d\tilde k\ L_{il}(x,t; \tilde k)\ {\mathbb M}_{lj}(x,t; \tilde k, k) =
\int d\tilde k\ \hat X_i(x,t; \tilde k)\ \hat P_{ij}(x,t; \tilde k, k)
\ee
the other quantities are then immediately deduced via (\ref{basic1}), (\ref{basic2}). Thus the fields can be completely reconstructed from the knowledge of the kernel $K_{ij}$.
Similar expressions are then obtained for $K_{11}$ and $K_{i1}$ via (\ref{basic1}), (\ref{basic2}):
\ba
&& K_{11}(x, z) = - \sum_j \ \int d\tilde k \int dk\ L_j(x,t;k)P_{jj}(x,t;k, \tilde k)\hat Z_j(z; \tilde k)\cr
&& K_{i1}(x,z)=-\int d\tilde k\ \hat X_i(x,t; \tilde k)\hat Z_i(z; \tilde k)-\sum_j \int dk \int d\tilde k\ L_{ij}(x,t;k)\ P_{jj}(x,t; k , \tilde k) \ \hat Z_j(z; \tilde k). \cr
&&
\ea

Let us finally discuss in more detail the time part of the BT. As explained in detail earlier in the text, as well as in previous related works we are mostly interested in the time evolution of the defect. In the present formulation the defect degrees of freedom are encoded in ${\cal K}$, therefore studying the time evolution of ${\cal K}$ is of great relevance in this context. This will naturally lead to BT type relations similar to the ones derived in the previous sections as will become apparent below. Let ${\mathbb G}$ be the global Darboux transformation such that:
\be
\hat \Psi = {\mathbb G}\ \Psi
\ee
and $\Psi,\ \tilde \Psi$ satisfy:
\ba
&& ia\ \partial_t \hat  \Psi = \hat {\cal D}\ \hat \Psi\cr
&& ia\ \partial_t \Psi = {\cal D}\  \Psi
\ea
where
\be
{\cal D} = {\cal D}_0 + M,~~~~\hat {\cal D} = \hat  {\cal D}_0 + \hat M. \label{A}
\ee
From the latter equations immediately follows the typical time part of a Darboux-B\"{a}cklund transformation
\be
\partial_t{\mathbb G}= \hat {\cal D}\ {\mathbb G} - {\mathbb G}\ {\cal D}. \label{B}
\ee
Taking into account (\ref{A}), (\ref{B}) and setting ${\mathbb G} = \mathds{1}+{\cal K}$,
we obtain the following global expression:
\be
i a\ \partial_t {\cal K} = \hat M\ {\cal K} - {\cal K}\ M + {\cal D}_0\ {\cal K} - {\cal K}\ {\cal D}_0 +\hat M - M
\ee
$M,\ \hat M$ are ${\cal N}\times {\cal N}$ matrices, and ${\cal D}_0 = \mathds{1} \partial_x^2$.
The integral representation of the expression above becomes ($K$ are also ${\cal N}\times {\cal N}$ matrices)
\ba
ia\ \int_{-\infty}^x dy\ \partial_t K(x,y) {\mathrm f}(y) &=& \int_{-\infty}^x dy\ M(x) K(x,y){\mathrm f}(y) - \int_{-\infty}^x dy\  K(x,y)M(y){\mathrm f}(y) \cr
&+&  \int_{-\infty}^x dy\ \Big (\partial_x^2 K(x,y) - \partial_y^2 K(x,y)\Big ){\mathrm f}(y) \cr
&+& 2 \partial_x K(x,x) {\mathrm f}(x) + (\hat M(x) - M(x)){\mathrm f}(x)
\ea
leading to the following equations:
\ba
&& ia\ \partial_t K(x,y) = \partial_x^2 K(x,y) - \partial_y^2 K(x,y) +\hat M(x)\ K(x,y) - K(x,y)\ M(y) \cr
&& 2 \partial_x K(x,x) = M(x) - \hat M(x). \label{TE}
\ea

One can of course start the ``dressing'' process with trivial solutions i.e. $M=0$ as described in the previous subsection. But in general the time evolution (\ref{TE}) describes the connection between two different solutions of the same non-linear differential equation. With this we conclude our analysis on the Darboux transforms and dressing for the vector NLS model.

\end{document}